\documentstyle[12pt,psfig]{article}
\begin{document}
\author{Thomas C. Bishop\footnote{Center for Bioenvironmental Research at Tulane and Xavier
Universities; \newline 1430 Tulane Avenue, New Orleans, LA 70112
Phone: (504)-988-6203,  Fax: (504)-585-6428, E-mail:
bishop@TCS.tulane.edu} \and Oleksandr O. Zhmudsky\footnote{Center
for Bioenvironmental Research at Tulane and Xavier
Universities\newline Phone: (504)-585-6145,  E-mail:
ozhmuds@tulane.edu}}
\date{Keywords: cell signaling, elastic rods, DNA, protein-protein
signaling, wave propagation\hspace{1mm}}
\title{Elastic wave propagation along DNA.}
\maketitle
\begin{abstract}
 It is shown that information transmission inside a cell can
occur by means of mechanical waves transmitted through DNA. The
propagation of the waves is strongly dependent on the shape of the
DNA  thus proteins that change the shape of DNA can  alter signal
transmission.  The overall effect is a method of signal processing
by DNA binding proteins that creates  a "cellular communications
network".

   The propagation of small amplitude disturbances through DNA is
treated according to the mechanical theory of elastic rods.
According to the theory four types of mechanical waves affecting
extension(compression), twist, bend or shear can propagate through
DNA.  Each type of wave has unique characteristic properties.
Disturbances affecting all wave types can propagate independently
of each other. Using a linear approximation to the theory of
motion of elastic rods, the dispersion of these waves is
investigated. The phase velocities of the waves lies in the range
$5 \sim 8 \r A/ps$ using constants suitable for a description of
DNA. The dispersion of all wave types of arbitrary wave length is
investigated for a straight, twisted rod.

   Based on these findings, we propose all-atom numerical simulations
of DNA to investigate the propagation of these waves as an
alternative measure of the wave velocity and dispersion analysis.

\end{abstract}
{\bf Introduction}

Our treatment of DNA is based on the hypothesis that such a well
organized system as a cell must have a highly sophisticated system
of communications. For instance, the cell cycle is the result of a
series of complex self organized processes that must be well
orchestrated. A model has been developed to demonstrate how these
events may be initiated by critical concentrations of specific
proteins which shift equilibrium to favor the advance of the cell
cycle\cite{John}. However, it is also known that the cell-cycle
can be disrupted if conditions are not suitable. Similar arguments
apply to virtually all cellular processes and in order to achieve
the required checks and balances a method of communication is
necessary.

Is there the possibility that information is transmitted through
DNA electromagnetically instead of mechanically? In such case DNA
will function as a transmission line which requires total internal
reflection (TIR) of the radiation within the DNA. To achieve TIR,
the wavelength of the radiation must be 5-10 times less than the
diameter of the transmission line. Since the diameter of the DNA
is close to 20 $\r{A}$ the radiation must have a wavelength close
to 2 $\r{A}$. This wavelength is close to atomic dimensions so
diffraction should dominate.  Furthermore, the energy associated
with this wavelength radiation is on the order of $10^5\ kcal/mol$
which is sufficient to destroy chemical bonds and therefore not
easily managed biologically. For these reasons, we believe that
the communication will be achieved mechanically.

The remainder of the paper will demonstrate that the mechanical
properties of DNA have the necessary
 time and spatial dimension to support the propagation of information
 while interactions between proteins and DNA provide a
mechanism by which this information is processed.

   This manuscript is based on a continuous medium model of DNA.
Such an approach is well known and widely used to describe solids,
liquids and gases (see, for example \cite{Landau1,Landau2}). Thus,
when we speak of infinitely small elements of volume, we shall
always mean those which are "physically" infinitely small, i.e.
very small compared with wavelength or radius of curvature under
consideration, but large compared with the distance between the
atoms in DNA.

\section{System of equation for the elastic rod dynamics}
 We utilize an elastic rod model described by \cite{SMK} that has been
 parameterized to represent DNA \cite{YMS-JEH}.  The equations of
 motion are as follows:

\begin{eqnarray}
\rho\left({\partial\vec\gamma\over\partial t}+\vec\omega\times
\vec\gamma \right)&=&\hat
C\cdot{\partial(\vec\Gamma-\vec\Gamma_0(s)) \over\partial s}+\vec
\Omega\times\left(\hat C\cdot\left(\vec\Gamma-\vec\Gamma_0\right)\right) \label{intro:1}\\
\hat I\cdot{\partial\vec\omega\over\partial t}+\vec\omega\times
\left(\hat I\cdot\vec\omega\right)&=&\hat
D\cdot{\partial(\vec\Omega-\vec\Omega_0(s)) \over\partial s}+\vec
\Omega\times\left(\hat
D\cdot\left(\vec\Omega-\vec\Omega_0\right)\right)\nonumber\\ &+&
\vec\Gamma\times\left(\hat
C\cdot\left(\vec\Gamma-\vec\Gamma_0\right)\right) \label{intro:2}\\
{\partial\vec\Gamma\over\partial t}+\vec\omega\times
\vec\Gamma&=&{\partial\vec\gamma\over\partial s}+\vec\Omega\times
\vec\gamma
\label{intro:3}\\
{\partial\vec\Omega\over\partial t}+\vec\omega\times
\vec\Omega&=&{\partial\vec\omega\over\partial s}
\label{intro:4}\end{eqnarray}

In this system of equations, equation (\ref{intro:1}) represents
the balance of force and linear momentum and equation
(\ref{intro:2}) represents the balance of torque and angular
momentum according to Newton's Laws. Equations (\ref{intro:3}) and
(\ref{intro:4}) are continuity relations expressed in a
non-inertial reference frame. Here, $s$ and $t$ are independent
variables representing time and fiducial arclength, respectively
\footnote{Fiducial arc length as describe in \cite{SMK} is the
same as actual arc length only when the rod is neither extended
nor compressed. During the propagation of a disturbance along the
rod, the rod will generally experience extension in some places
and compression in others so the actual arc length will generally
be different from fiducial arc length. Actual arc length plays no
role in the equations of motion}. The functions that we wish to
study  are the four three-vector functions $\vec\Omega$,
$\vec\Gamma$, $\vec\omega$, and $\vec\gamma$.

The matrices $\hat I$, $\hat C$ and $\hat D$ and the scalar $\rho$
are as follows: Matrix $\hat I$ is the linear density of the
moment of inertia tensor. Matrices $\hat C$ and $\hat D$ represent
the elastic properties of the rod ($C_3$: Young's modulus, $C_1$
and  $C_2$: shear modulus, $D_3$: torsional rigidity, $D_1$ and
$D_2$: bend stiffness).  $\rho$ is the linear mass density and is
equal to $3.22\times 10^{-15}\mbox{K/M}$ for an isotropic model of
DNA. The matrices have the following values for an isotropic model
of DNA:

\begin{eqnarray}
I_{ik}&=&\left(\matrix{I & 0 & 0 \cr
 0 & I & 0 \cr
 0 & 0  & 2I \cr}\right)\equiv
 \left(\matrix{4.03 & 0 & 0 \cr
 0 & 4.03  & 0 \cr
 0 & 0  & 8.06 \cr}\right)\times10^{-34}\left[KM\right]
\label{intro:6}\\
C_{ik}&=&\left(\matrix{C_1 & 0 & 0 \cr
 0 & C_2 & 0 \cr
 0 & 0  & C_3 \cr}\right)\equiv
 \left(\matrix{8.16 & 0 & 0 \cr
 0 & 8.16  & 0 \cr
 0 & 0  & 21.6 \cr}\right)\times10^{-10}\left[\frac {KM}{S^2}\right]
\label{intro:7}\\
D_{ik}&=&\left(\matrix{D_1 & 0 & 0 \cr
 0 & D_2 & 0 \cr
 0 & 0  & D_3 \cr}\right)\equiv
 \left(\matrix{2.7 & 0 & 0 \cr
 0 & 2.7  & 0 \cr
 0 & 0  & 2.06 \cr}\right)\times10^{-28}\left[\frac {KM^3}{S^2}\right]
\label{intro:8}\end{eqnarray}
  For isotropic bending $C_1=C_2$  and shear $D_1=D_2$ we will denote $C=C_1=C_2$ and
$D=D_1=D_2$.
  These definitions will be used below. The above values have been
adapted from \cite{Moroz}, \cite{Bouchiat}.

In biological terminology the components of $\vec\Gamma$
correspond to the three DNA helical parameters describing
translation $\vec\Gamma=(\mbox{shift,slide,rise})$ and the
components of $\vec\Omega$ correspond to the three DNA helical
parameters describing rotation
$\vec\Omega=(\mbox{tilt,roll,twist})$. If we attach a local
coordinate frame, denoted by $\{\vec d_1,\vec d_2,\vec d_3\}$, to
each base-pair the axes will point in the direction of the major
groove, minor groove and along the DNA helical axis, respectively.
The shape of the DNA can be described by a three-dimensional
vector function $\vec r(s,t)$ which is related to $\vec\Gamma$ and
$\vec\Omega$ by a suitable mathematical integration.

In elastic rod terminology the three-dimensional vector function
$\vec r(s,t)$ gives the centerline curve of the rod as a function
of $s$ and $t$, showing only
 how the rod bends. To show the twist, shear and extension
of the rod we must attach "director" frames made of the orthogonal
triples $\{\vec d_1,\vec d_2,\vec d_3\}$  at regular intervals
along the rod. The director frames are evenly spaced along the rod
when the rod is not extended  and are all parallel to each other
(with $\vec d_3$ pointing along the rod) when the rod is not bent
or twisted.  Any deformation of the rod will be indicated by a
corresponding change in the orientation of the director frames.

The vector $\vec\gamma$ ($\vec\Gamma$) is the translational
velocity of $\vec r$ in time (space):
\[\vec\gamma={\partial\vec r\over\partial t} \qquad
\vec\Gamma={\partial\vec r\over\partial s}\]
 The directors are of constant unit
length, so the velocity of each director $\vec\omega$
($\vec\Omega$) is always perpendicular to itself, and it must also
be perpendicular to the axis of rotation of the local frame.
\[{\partial\vec d_k\over\partial t}=\vec\omega\times\vec d_k\qquad
{\partial\vec d_k\over\partial s}=\vec\Omega\times\vec d_k\]

We will denote $\vec\Gamma$ of the unstrained state by
$\vec\Gamma_0$. In general an unstrained rod can have any shape.
The most simple case is when the rod is straight with unit
extension, $\vec\Gamma_0=\{0,0,1\}$, because in the unstrained
state:
$${\partial\vec r_3\over\partial s}=\{0,0,1\}$$

Similarly, an unstressed elastic rod may have an intrinsic bend
and/or twist, denoted by $\vec\Omega_0$, which must be subtracted
from the total $\vec\Omega$ to give the dynamically active part.
The simplest case is a rod with no intrinsic bend or twist
$\vec\Omega=\{0,0,0\}$ and $\vec\Gamma_0=\{0,0,1\}$.

For  short wave lengths (Section \ref{section:2}) we will suppose
that $\vec\Omega_0=$ \linebreak
$(\Omega_{01},\Omega_{02},\Omega_{03})=const$ and
$\vec\Gamma_0=(\Gamma_{01},\Gamma_{02},\Gamma_{03})=const$. In
section \ref{section:3} we will consider a straight rod with an
intrinsic twist $\vec\Omega=\{0,0,\Omega_0\}$ and
$\vec\Gamma=\{0,0,\Gamma_0\}$.

\section{System for small amplitude waves}\label{section:2}

The results from this section are valid for any shape in which the
curvature of the rod is much greater than the wavelength of the
disturbance being propagated regardless of the intrinsic shape of
the rod.

Let us search as usual for the equilibrium point of the system
(\ref{intro:1}-\ref{intro:4}) by setting $\vec\gamma$,
$\vec\Gamma$, $\vec\omega$ and $\vec\Omega$ to constants. It is
easy to see that an equilibrium point of equations
(\ref{intro:1}-\ref{intro:4}) is $\vec\gamma=\vec\omega=0$,
$\vec\Gamma=\vec\Gamma_0$ and $\vec\Omega=\vec\Omega_0\;$

We shall suppose that each variable differs slightly from the
equilibrium value so we  retain only the linear terms in the
equations. The linearized system (\ref{intro:1}-\ref{intro:4}) is:
\begin{eqnarray}
\rho{\partial\vec\gamma\over\partial t} &=& \hat
C\cdot{\partial\vec\Gamma\over\partial s}+\vec
\Omega_0\times\left(\hat C\cdot\vec\Gamma\right) \label{linear:1}\\
\hat I\cdot{\partial\vec\omega\over\partial t} &=& \hat
D\cdot{\partial\vec\Omega \over\partial s}+\vec
\Omega_0\times\left(\hat D\cdot\vec\Omega\right)+
\vec\Gamma_0\times\left(\hat
C\cdot\vec\Gamma\right) \label{linear:2}\\
{\partial\vec\Gamma\over\partial t} &+&
\vec\omega\times\vec\Gamma_0 =
{\partial\vec\gamma\over\partial s} + \vec\Omega_0\times\vec\gamma\label{linear:3}\\
{\partial\vec\Omega\over\partial t} &+&
\vec\omega\times\vec\Omega_0 = {\partial\vec\omega\over\partial s}
\label{linear:4}\end{eqnarray}
  In the system (\ref{linear:1}-\ref{linear:4}) and further in this
paper  $\vec\gamma$, $\vec\Gamma$, $\vec\omega$ and $\vec\Omega$
denote the small deviations from the equilibrium values, e.g.
$\vec\Gamma-\vec\Gamma_0\to\vec\Gamma$.

In the usual manner we search for a solution to
(\ref{linear:1}-\ref{linear:4}) using harmonic analysis. We assume
that each variable depends on time and space as follows:
\begin{equation}\label{linear:4a}
\vec G_i(s,t)=\vec G_{i0}\cdot e^{\displaystyle {-i\omega t +
iks}}
\end{equation}
where $\vec G_i(s,t)$ denotes one of the four vector variables
from the system of equations (\ref{linear:1}-\ref{linear:4}) and
$\vec G_{i0}=const$ denotes its amplitude.\footnote{Note that
$\omega$ is a scalar quantity corresponding to a frequency of
oscillation and that $\vec\omega$ is a vector quantity that
describes a rotation, so there should be no confusion between the
two variables.}

Let us substitute (\ref{linear:4a}) to the system
(\ref{linear:1}-\ref{linear:4}):
\begin{eqnarray}
-i\omega\rho\vec\gamma &=& ik(\hat C\cdot\vec\Gamma)+ \vec
\Omega_0\times\left(\hat C\cdot\vec\Gamma\right)
 \label{linear:4b}\\
-i\omega (\hat I\cdot\vec\omega) &=& ik\left(\hat
D\cdot\vec\Omega\right)+ \vec\Omega_0\times\left(\hat
D\cdot\vec\Omega\right)+ \vec\Gamma_0\times\left(\hat
C\cdot\vec\Gamma\right)\label{linear:4c}\\
-i\omega\vec\Gamma + \vec\omega\times\vec\Gamma_0 &=&
ik\vec\gamma + \vec\Omega_0\times\vec\gamma\label{linear:4d}\\
-i\omega\vec\Omega + \vec\omega\times\vec\Omega_0&=& ik\vec\omega
\label{linear:4e}
\end{eqnarray}

  For short wavelength we search for solutions with
$\omega\propto k$, so in the limit $k\to\infty$ we neglect all
terms which contain cross products and the system
(\ref{linear:4b}-\ref{linear:4e}) becomes:
\begin{eqnarray}
(\omega^2\rho - k^2 C)\vec\Gamma_\bot &=& 0
\label{linear:4k}\\
(\omega^2\rho - k^2 C_3)\Gamma_3 &=& 0
\label{linear:4l}\\
(\omega^2 I - k^2 D)\vec\Omega_\bot &=& 0 \label{linear:4m} \\
(2\omega^2 I - k^2 D_3)\Omega_3 &=& 0 \label{linear:4n}
\end{eqnarray}
  Here division into longitudinal and transversal components has
  been obtained by defining $\vec\Gamma_\bot\equiv(\Gamma_1,\Gamma_2,0)$ and
$\vec\Omega_\bot\equiv(\Omega_1,\Omega_2,0)$.

  Equations (\ref{linear:4k}-\ref{linear:4n}) represent four
types of waves that can propagate in the elastic rod with
velocity, $V$ (according to the order of equations above):
\begin{itemize}
  \item Shear waves ($\vec\Gamma_\bot$):
\begin{equation}\label{linear:4p}
  V_{Shear}=\sqrt{\frac {C}\rho}=\sqrt{8.16\cdot 10^{-10}
[KM/s^2]\over 3.22 \cdot 10^{-15} [K/M]}\approx 5.03 \r A/ps
\end{equation}
  \item Extension waves ($\Gamma_3$):
\begin{equation}\label{linear:4q}
  V_{Extension}=\sqrt{\frac {C_3}\rho}=\sqrt{21.6\cdot 10^{-10}
[KM/s^2]\over 3.22 \cdot 10^{-15} [K/M]}\approx 8.2 \r A/ps
\end{equation}
  \item Bend waves ($\vec\Omega_\bot$):
\begin{equation}\label{linear:4r}
V_{Bend}=\sqrt{\frac DI} =\sqrt{2.7\cdot 10^{-28} [KM^3/s^2]\over
4.03 \cdot 10^{-34} [KM]}\approx 8.2 \r A/ps
\end{equation}
  \item Twist waves ($\Omega_3$):
\begin{equation}\label{linear:4s}
V_{Twist}=\sqrt{\frac {D_3}{2I}} =\sqrt{2.06\cdot 10^{-28}
[KM^3/s^2]\over 2\cdot 4.03 \cdot 10^{-34} [KM]}\approx 5.0 \r
A/ps
\end{equation}
\end{itemize}

These results were obtained for a rod  with arbitrary shape
because all terms that define the rod shape in equations
(\ref{linear:4b}-\ref{linear:4e}) were omitted. So, if wavelength
tends to zero\footnote{We must note once more that "wavelength
tends to zero" means that  it is still much greater than distance
between atoms, see Introduction).} (is the least space parameter
of the problem) four wave types can propagate along the rod.

It is well known that the measurement of wave velocity is a usual
method for determining  elastic properties of solids. So, wave
velocity (transversal and longitudinal) and Young's modulus (shear
modulus) are uniquely determined  by formulae similar to the
(\ref{linear:4p}-\ref{linear:4s}).

Hakim et al. \cite{Hakim} have determined the velocity of sound in
DNA. Based on elastic constants obtained from \cite{Moroz},
\cite{Bouchiat} and used in this work, the velocity  is
approximately two times less than that obtained by Hakim et al.
\cite{Hakim}.

Experiments to measure elastic properties of single DNA molecules
have been reported using scanning tunnelling microscopy
\cite{Guckenberger}, fluorescence microscopy \cite{Yanagida},
fluorescence correlation spectroscopy \cite{Wennmalm}, optical
tweezers \cite{Bustamante1}\cite{Wang}, bead techniques in
magnetic fields \cite{Bustamante2} \cite{Strick}, optical
microfibers \cite{Cluzel}, low energy electron point sources
(electron holography) \cite{Fink} and atomic force microscopy
(AFM) \cite{Hansma}.  Each method differs in the molecular
properties probed,  spatial and temporal resolution, molecular
sensitivity and working environment, so each method gives close
but not the same value of elastic constants. So if we evaluate
extension/compression velocity using constants obtained from the
data reported in \cite{Bustamante1} the result will be two times
slower than the value reported here and based on data from
\cite{Moroz} and \cite{Bouchiat}.

\section{Linear waves in the straight rod with no
intrinsic bend or shear.}\label{section:3}
  Here we  concentrate our attention  on the
case of the straight, twisted rod with no intrinsic bend or shear.
In this section we suppose that $\vec\Omega_0=(0,0,\Omega_0)$ and
$\vec\Gamma_0=(0,0,\Gamma_0)$. As we will see this case allows for
a complete analytical analysis of extension/compression (Section
\ref{Sound}), twist (Section \ref{Twist}), and  bend/shear
(Section \ref{Coupled}) waves of arbitrary wave length. Equations
(\ref{linear:1}-\ref{linear:4}) in component form are:

\begin{eqnarray}
\rho{\partial\gamma_1\over\partial t} &=&
C\cdot{\partial\Gamma_1\over\partial s}-C\Omega_0\Gamma_2
 \label{linear:9}\\
\rho{\partial\gamma_2\over\partial t} &=&
C\cdot{\partial\Gamma_2\over\partial s}+C\Omega_0\Gamma_1
\label{linear:10}\\
\rho{\partial\gamma_3\over\partial t} &=&
C_3\cdot{\partial\Gamma_3\over\partial s}\label{linear:11}\\
I\cdot{\partial\omega_1\over\partial t} &=&
D\cdot{\partial\Omega_1\over\partial s}-D\Omega_0\Omega_2-
C\Gamma_0\Gamma_2 \label{linear:12}\\
I\cdot{\partial\omega_2\over\partial t} &=&
D\cdot{\partial\Omega_2\over\partial s}+D\Omega_0 \Omega_1+
C\Gamma_0\Gamma_1 \label{linear:13}\\
2I\cdot{\partial\omega_3\over\partial t} &=&
D_3\cdot{\partial\Omega_3\over\partial s}\label{linear:14}\\
{\partial\Gamma_1\over\partial t} + \Gamma_0\omega_2 &=&
{\partial\gamma_1\over\partial s} - \Omega_0\gamma_2\label{linear:15}\\
{\partial\Gamma_2\over\partial t} - \Gamma_0\omega_1 &=&
{\partial\gamma_2\over\partial s} + \Omega_0\gamma_1\label{linear:16}\\
{\partial\Gamma_3\over\partial t} &=&
{\partial\gamma_3\over\partial s}\label{linear:17}\\
{\partial\Omega_1\over\partial t} + \Omega_0\omega_2&=&
{\partial\omega_1\over\partial s} \label{linear:18}\\
{\partial\Omega_2\over\partial t} - \Omega_0\omega_1 &=&
{\partial\omega_2\over\partial s} \label{linear:19}\\
{\partial\Omega_3\over\partial t} &=&
{\partial\omega_3\over\partial s} \label{linear:20}
\end{eqnarray}

It is easy to see that equations (\ref{linear:11}) and
(\ref{linear:17}) (also (\ref{linear:14}) and (\ref{linear:20}))
are independent from all other equations and describe extension
(twist) waves. These two wave types will be discussed in the next
two sections.

\subsection{Sound (extension/compression) waves}\label{Sound}
  From equations (\ref{linear:11}) and
(\ref{linear:17}) it is easy to obtain two wave equations for the
extension/compression waves:
\begin{eqnarray}
\rho{\partial^2\gamma_3\over\partial t^2}&-&
C_3{\partial^2\gamma_3\over\partial s^2}=0 \label{Sec1:9}\\
\rho{\partial^2\Gamma_3\over\partial t^2}&-&
C_3{\partial^2\Gamma_3 \over\partial s^2}=0 \label{Sec1:10}
\end{eqnarray}
 These small amplitude waves have velocity (\ref{linear:4q})
and these equations have an harmonic solution\footnote{Equations
(\ref{Sec1:9}-\ref{Sec1:10}) have not only harmonic solutions. In
general  the solutions will be (d'Alembertian waves):
\begin{eqnarray}
\gamma_3=p_1(s-t\sqrt{\frac {C_3}\rho})+p_2(s+t\sqrt{\frac
{C_3}\rho})\label{Sec1:11} \\
\Gamma_3=q_1(s-t\sqrt{\frac {C_3}\rho})+q_2(s+t\sqrt{\frac
{C_3}\rho})\label{Sec1:12}
\end{eqnarray}

 where $p_{1,2}$ and $q_{1,2}$ are arbitrary functions and $\dot p_{1,2}=\mp\sqrt{C^3/\rho}\;\dot q_{1,2}$. The
point to keep in mind is that the amplitude of the sound waves
must be small. It is easy to see the movement follows directly
from  equations (\ref{Sec1:11}) or (\ref{Sec1:12}). It is clear
that the arbitrary function $p_1$ has the same value for all
arguments $s-t\sqrt{{C_3}/\rho}=const$. This relationship
describes a straight line in the $(s,t)$-plane so $p_1$ is
constant along this line. The same statement holds for all
neighboring points so an initial shape moves along the
$s-t\sqrt{{C_3}/\rho}=const$ line with velocity $\sqrt{ {C_3}
/\rho}$. The arbitrary function $p_2$ describes motion in the
opposite direction. }:

\begin{eqnarray}
\Gamma_3=A_1\sin(s-t\sqrt{\frac {C_3}\rho})+A_2\cos(s-t\sqrt{\frac
{C_3}\rho})\label{harm:11} \\
\gamma_3=B_1\sin(s-t\sqrt{\frac {C_3}\rho})+B_2\cos(s-t\sqrt{\frac
{C_3}\rho})\label{harm:12}
\end{eqnarray}
  where $A_{1,2}$ are  arbitrary constants (wave
amplitudes) and $B_{1,2}=\phantom{=}$ $-\sqrt{C_3/\rho}\;A_{1,2}$.

 Using elastic constants obtained from  data in \cite{Moroz} and \cite{Bouchiat}
 the velocity of sound in DNA is equal to
$V_{sound}= 8.2 \r A/ps $ (see equation (\ref{linear:4q})) and
dispersion is linear. This is the velocity of propagation of
harmonic waves in DNA according to the linear approximation. In
the linear approximation the amplitude of the wave must be small,
but it can have any wavelength.

\subsection{Twist waves}\label{Twist}
In the same manner equations (\ref{linear:14}) and
(\ref{linear:20}) yield two  wave equations which describe  twist
waves:

\begin{eqnarray}
2I{\partial^2\omega_3\over\partial t^2}&-&
D_3{\partial^2\omega_3\over\partial s^2}=0 \label{TwistPure:6}\\
2I{\partial^2\Omega_3\over\partial t^2}&-& D_3{\partial^2\Omega_3
\over\partial s^2}=0 \label{TwistPure:7}
\end{eqnarray}
  Linear solutions of these equations can be written as:
\begin{eqnarray}
\Omega_3=P_1 \sin(s-t\sqrt{\frac {D_3}{2I}})+P_2
\cos(s-t\sqrt{\frac
{D_3}{2I}})\label{TwistPure:8} \\
\omega_3=Q_1 \sin(s-t\sqrt{\frac {D_3}{2I}})+Q_2
\cos(s-t\sqrt{\frac {D_3}{2I}})\label{TwistPure:9}
\end{eqnarray}
  where $P_{1,2}$ are the arbitrary constants (wave
amplitudes)  and $Q_{1,2}=\phantom{=}$
$-\sqrt{{D_3}/{2I}}\;P_{1,2}$.

 These solutions describe twist waves with velocity
$V_{twist}=\sqrt{\displaystyle {D_3}/{2I}}\approx 5.0 \r A/ps$
(see equation (\ref{linear:4s})). In the general case, the
solutions of these equations will be d'Alembertian waves described
by (\ref{Sec1:11}-\ref{Sec1:12}). Again the dispersion law is
linear.

\subsection{Bend and Shear Waves}\label{Coupled}

Substituting (\ref{linear:4a}) into the remaining equations of
system (\ref{linear:9}-\ref{linear:20}) yield:
\begin{eqnarray}
-i\omega\rho\vec\gamma_\bot &=& ikC\vec\Gamma_\bot+
C\left[\vec\Omega_0\times\vec\Gamma_\bot\right]
 \label{linear:42}\\
-i\omega I\vec\omega_\bot &=& ikD\vec\Omega_\bot+
D\left[\vec\Omega_0\times\vec\Omega_\bot\right]+
C\left[\vec\Gamma_0\times\vec\Gamma_\bot\right]\label{linear:43}\\
-i\omega\vec\Gamma_\bot +
\left[\vec\omega_\bot\times\vec\Gamma_0\right] &=&
ik\vec\gamma_\bot + \left[\vec\Omega_0\times\vec\gamma_\bot\right]\label{linear:44}\\
-i\omega\vec\Omega_\bot +
\left[\vec\omega_\bot\times\vec\Omega_0\right]&=&
ik\vec\omega_\bot \label{linear:45}
\end{eqnarray}

Here, the definitions
$\vec\gamma_\bot\equiv(\gamma_1,\gamma_2,0)$,
$\vec\Gamma_\bot\equiv(\Gamma_1,\Gamma_2,0)$,
$\vec\omega_\bot\equiv(\omega_1,\omega_2,0)$ and
$\vec\Omega_\bot\equiv(\Omega_1,\Omega_2,0)$ are used. This is an
homogeneous system of linear equations with unknowns
$\vec\gamma_\bot$, $\vec\Gamma_\bot$, $\vec\omega_\bot$ and
$\vec\Omega_\bot$. Solutions other than the trivial solution exist
only if the determinant of the coefficients is zero. This
condition is satisfied if:
\begin{eqnarray}
\left(\omega^2\rho-(k^2+\Omega_0^2)C-{\rho\omega^2
C\Gamma_0^2\left(\omega^2I -(k^2+\Omega_0^2)D\right)
\over\left(\omega^2I-(k^2
+\Omega_0^2)D\right)^2-4k^2D^2\Omega_0^4}\right)^2\nonumber \\
=\left(2kC\Omega_0+{2k\rho D\omega^2 C\Gamma_0
\left(\vec\Omega_0\cdot\vec\Gamma_0\right)\over
\left(\omega^2I-(k^2+\Omega_0^2)D\right)^2-4k^2D^2\Omega_0^4}\right)^2
\label{linear:46}\end{eqnarray}

Let us discuss the untwisted rod ($\Omega_0=0$)\footnote{For
untwisted rod system (\ref{linear:9}-\ref{linear:20}) consists of
two independent subsystems. Equations (\ref{linear:9},
\ref{linear:13}, \ref{linear:15}, \ref{linear:19}) and
(\ref{linear:10}, \ref{linear:12}, \ref{linear:16},
\ref{linear:18}) present two different polarizations of bend/shear
wave.}. Equation (\ref{linear:46}) yields:
\begin{equation}\label{linear:47}
\omega^2\rho-k^2 C-{\rho\omega^2 C\Gamma_0^2\over \omega^2 I-k^2
D}=0
\end{equation}

  It will be convenient to use dimensionless variables (marked by
asterisk) $\omega^*=\omega\sqrt{I/C}$, $k^*=k\sqrt{I/\rho}$ in
this section and define $G^2\equiv {D/I\over C/\rho}$. Equation
(\ref{linear:47}) becomes(asterisks are omitted):
\begin{equation}\label{linear:48}
(\omega^2-k^2)(\omega^2-G^2k^2)=\Gamma_0^2\omega^2,
\end{equation}
with solutions:
\begin{equation}\label{linear:48a}
\omega_{1,2,3,4}= \pm \frac 12\left\{ \sqrt{[G+1]^2k^2+\Gamma^2_0}
\pm \sqrt{[G-1]^2k^2+\Gamma^2_0} \right\}
\end{equation}

\begin{figure*}[ht]
\caption{Dispersion of Bend/Shear Waves in DNA.}\label{figure:1}
\begin{center}
\psfig{figure=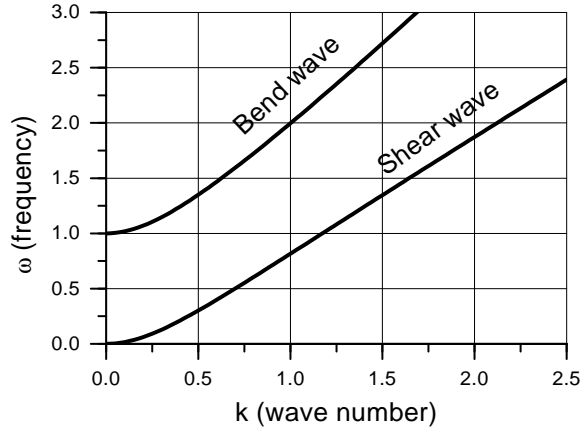,width=90mm}
\end{center}
\end{figure*}

There are a total of four solutions.  The $\pm$ sign before the
braces merely changes the direction of propagation and will not be
discussed further. The $\pm$ sign between the roots produces two
different branches of solutions as presented in
Fig.\ref{figure:1}. The "upper" branch\footnote{In this way we
will name branch that does not go to zero in the limit $k\to 0$.}
corresponds to  bend waves and the "lower" branch to  shear waves.

For a rod of infinite length, any value of $\omega$ can be chosen
that satisfies equation (\ref{linear:48}), but for a rod of finite
length there will be an additional constraint that must be
satisfied. If the rod is fixed at both ends then the usual
restriction of having a node at each end applies. This introduces
the constraint $L=(n/2)\lambda$ between the length of the rod,
$L$, and the possible wavelengths, $\lambda$, so that only
discrete points along either sub-branch in Figure 1 will be
observed.

\section{Conclusion.}
The primary results are that, in the linear approximation, four
different types of waves can propagate through a uniform elastic
rod. These waves correspond to extension(compression), twist, bend
and shear. An extension or twist wave will propagate without
exciting other modes or changing shape, and it has a linear
dispersion relation. Bend and shear waves behave rather
differently and have a nonlinear dispersion relation. Each type
obeys a dispersion law that describes two additional sub-branches.

Utilizing constants suitable for DNA we find that, in the limit of
small wavelengths, extension and bend propagate with a velocity of
approximately 8\AA/ps and twist and shear propagate with a
velocity of approximately 5\AA/ps.

Is is also significant that the dispersion relation for bend and
shear waves is coupled with the inherent twist of DNA. We propose
that these physical phenomena enable proteins interacting with DNA
to accomplish highly sophisticated tasks. For example the
difference in extension and twist velocities can be utilized to
measure the distance between two points on the DNA. Since the
dispersion law for bend and shear waves depends on intrinsic
twist, a mechanism for measuring DNA topology exists because of
the relation between twist, linking number and writhe. Other
protein-protein communications can certainly be established to
assist cellular mechanisms.

\subsection{Suggestion for a simulation experiment}
We suggest a molecular dynamics simulation experiment to check the
 correspondence between the linear theory of the propagation of waves in elastic
 rods and an all atom simulation of DNA.
 In all-atom molecular dynamics simulations of  DNA
the number of base pairs that can be simulated for a significant
length of time
 is from tens to less than hundreds of base pairs.
 For a simulation of DNA with fixed ends it
is sufficient to apply a sharp impulse ($\delta$ function
excitation) to one end of the DNA and measure the time of
propagation of this disturbance to the other end as a measure of
the velocity of extension/compression waves in DNA according to
equation (\ref{linear:4q}).  The propagation of twist can be
similarly tested by applying a torque.

To measure  dispersion one must make some simple calculations
before the simulation and then excite one end of the DNA with a
driving force of the appropriate frequency. In this manner a
standing wave can be established in the DNA. The wave frequency
and wave number ($2\pi /\lambda$) are related by the well known
relation:
\begin{equation}\label{length:1}
  \omega = V_{Wave} k
\end{equation}
 where $\omega=2\pi f$ is the wave frequency, $V_{Wave}$ is the wave velocity
and $k=2\pi/\lambda$ is the wavelength. In this case the wave
length (wave number) is easy to evaluate:
\begin{equation}\label{length:2}
  L = n{\lambda\over 2}, \qquad \mbox{where n=1,2,3,.. }
\end{equation}
 and $f$ is the frequency of the driving oscillation.
 The values obtained from simulation can be compared
 with the number obtained from formula
(\ref{linear:48} or \ref{linear:48}) as a means of correcting our
choice of constants.

\section{Acknowledgement}
    We express our thanks to Dr. Jan-Ake Gustafsson, Dr. Iosif
Vaisman and Dr. Yaoming Shi for stimulating and helpful
discussions. We would also like to thank Yaoming Shi for a
critical reading of the manuscript.

This work was conducted in the Theoretical Molecular Biology
Laboratory which was established under NSF Cooperative Agreement
Number OSR-9550481/LEQSF (1996-1998)-SI-JFAP-04 and supported by
the Office of Naval Research (NO-0014-99-1-0763).

\end{document}